\documentclass{osa-article}
\journal{boe}
\usepackage{xspace}
\usepackage{xcolor, graphicx, setspace, tocloft, here}

\newcommand{\invitro}{\textit{in vitro}\xspace}

\newcommand{\exvivo}{\textit{ex vivo}\xspace}

\newcommand{\enface}{\textit{en face}\xspace}
\newcommand{\Enface}{\textit{En face}\xspace}
\newcommand{\um}{\(\muup\)m\xspace}

\newcommand{\fig}{Fig.\@\xspace}

\newcommand{\etal}{\textit{et al.\@}\xspace}
\newcommand{\ave}[2]{\left<{#1}\right>_{#2}\xspace}

\begin{document}
\noindent This manuscript has been published in Biomedical Optics Express 13, 4071-4086 (2022).
\url{https://doi.org/10.1364/BOE.461433}.
\vspace{2ex}
	
\title{Label-free metabolic imaging of non-alcoholic-fatty-liver-disease (NAFLD) liver by volumetric dynamic optical coherence tomography}

\author{Pradipta Mukherjee,\authormark{1} Shinichi Fukuda,\authormark{2,3} Donny Lukmanto,\authormark{3} Toshiharu Yamashita,\authormark{4} Kosuke Okada,\authormark{5} Shuichi Makita,\authormark{1} Ibrahim Abd El-Sadek,\authormark{1, 6} Arata Miyazawa,\authormark{7} Lida Zhu,\authormark{1}, Rion Morishita,\authormark{1} Antonia Lichtenegger,\authormark{8, 1} Tetsuro Oshika,\authormark{2} and Yoshiaki Yasuno\authormark{1,*}}

\address{\authormark{1}Computational Optics Group, University of Tsukuba, Tsukuba, Ibaraki, Japan\\
\authormark{2}Department of Ophthalmology, Faculty of Medicine, University of Tsukuba, Tsukuba, Ibaraki, Japan\\
\authormark{3}Department of Advanced Vision Science, Faculty of Medicine, University of Tsukuba, Tsukuba, Ibaraki, Japan\\
\authormark{4}Laboratory of Regenerative Medicine and Stem Cell Biology, Graduate School of Comprehensive Human Sciences, University of Tsukuba, Tsukuba, Ibaraki, Japan\\
\authormark{5}Division of Medical Sciences, Faculty of Medicine, University of Tsukuba, Tsukuba, Ibaraki, Japan\\
\authormark{6}Department of Physics, Faculty of Science, Damietta University, 34517 New Damietta City, Damietta, Egypt\\
\authormark{7}Sky Technology Inc., Tsukuba, Ibaraki, Japan\\
\authormark{8}Center for Medical Physics and Biomedical Engineering, Medical University of Vienna, Austria}

\email{\authormark{*}Corresponding author: yasuno@optlab2.bk.tsukuba.ac.jp} 

\homepage{http://optics.bk.tsukuba.ac.jp/COG/} 


\begin{abstract}
Label-free metabolic imaging of non-alcoholic fatty liver disease (NAFLD) mouse liver is demonstrated \exvivo by dynamic optical coherence tomography (OCT).
The NAFLD mouse is a methionine choline-deficient (MCD)-diet model, and two mice fed MCD diet for 1 and 2 weeks are involved in addition to a normal-diet mouse.
The dynamic OCT is based on repeating raster scan and logarithmic intensity variance (LIV) analysis which enables volumetric metabolic imaging with a standard-speed (50,000 A-lines/s) OCT system.
Metabolic domains associated with lipid droplet accumulation and inflammation are clearly visualized three-dimensionally.
Particularly, the normal-diet liver exhibits highly metabolic vessel-like structures of peri-vascular hepatic zones.
The 1-week MCD-diet liver shows ring-shaped highly metabolic structures formed with lipid droplets.
The 2-week MCD-diet liver exhibits fragmented vessel-like structures associated with inflammation.
These results imply that volumetric LIV imaging is useful for visualizing and assessing NAFLD abnormalities.
\end{abstract}

\section{Introduction}
\label{sec: Intro}

Non-alcoholic fatty liver disease (NAFLD), the most common chronic liver disease, is found in populations with obesity or type 2 diabetes \cite{fabbrini_obesity_2010,el-serag_diabetes_2004}.
It also frequently occurs in healthy people without any obvious risk factors \cite{friedman_mechanisms_2018}. 
NAFLD is a wide spectrum of metabolic diseases \cite{petta_non-alcoholic_2009}.
The initial stage of NAFLD is hepatic steatosis \cite{angulo_nonalcoholic_2002} and then, it can progress to non-alcoholic steatohepatitis (NASH) \cite{marra_molecular_2008}, hepatic fibrosis \cite{schuster_triggering_2018}, cirrhosis \cite{kitade_crosstalk_2009}, and ultimately hepatocellular carcinoma (HCC) \cite{kanwal_risk_2018,yasui_characteristics_2011}.
After the transition from NASH to a more advanced stage of hepatic fibrosis, the disease becomes irreversible because of the collapse of the hepatic lobule and its substitution with fibrotic tissue\cite{popper_hepatic_1970, bataller_liver_2005}. 

Numerous studies have been carried out to understand the NAFLD pathogenesis and to develop novel therapeutic strategies.
However, NAFLD pathogenesis is not yet clearly understood, and no drug is available.
The only treatment is lifestyle modification by diet and exercise \cite{banini_nonalcoholic_2016}.
The use of small animal models has proven effective for the study of this disease not only for studying the pathogenesis but also for examining the therapeutic effects of various drugs \cite{anstee_mouse_2006, takahashi_animal_2012}.
Such animal models can correctly reflect both the histopathology and the pathophysiology of human NAFLD \cite{takahashi_animal_2012}.

According to animal-model-based studies, significant amount of lipid droplets (LDs) accumulate in the hepatocytes of hepatic steatosis \cite{kumar_natarajan_structure_2017,mashek_hepatic_2021}.
LDs provide the lipid source for energy metabolism.  
NASH is an inflammatory disease characterized by liver inflammation and ballooning of the hepatocytes \cite{sheka_nonalcoholic_2020}.
Several inflammatory cytokines are produced by inflammatory cells during NASH. 
Such inflammatory cytokines play a key role in the progression from hepatic steatosis to NASH \cite{silvana_inflammatory_2018, lopetuso_harmful_2018}. 
Therefore, during an early stage of NAFLD, LDs and inflammatory cells are accumulated and they are associated with high metabolism. 

To understand NAFLD pathogenesis, it is important to access the liver tissue metabolism including those associated with LD and inflammation.
In addition, since the LD accumulation and inflammation can occur not only at the tissue surface but also in deep tissue, it is important to visualize the metabolism of deep tissues without slicing the sample.

Assessment of NAFLD progression can be performed via conventional histological imaging of liver biopsy specimens. 
However, this method is invasive, destructive, time-consuming, and labor-intensive \cite{kleiner_design_2005}. 
Traditional non-invasive imaging modalities, including ultrasonography \cite{saverymuttu_ultrasound_1986}, computed tomography \cite{ricci_noninvasive_1997}, and magnetic resonance imaging \cite{szczepaniak_measurement_1999} can detect the presence of fat content in the liver.
However, such methods possess several limitations, including no sensitivity to tissue metabolism and low spatial resolution in the order of millimeters.
Fluorescence microscopy can also visualize the LDs and inflammatory cells \cite{fukumoto_deformation_2002, listenberger_fluorescent_2007, wang_altered_2010, somwar_live-cell_2011}, but it has two major limitations.
First, it requires the use of an exogenous contrast agent. 
Introducing such an agent may perturb the physiological condition of the tissue. 
Second, it has limited image penetration of a few hundred micrometers; hence, it is not a perfect modality for three-dimensional (3D) metabolism investigation.

Optical coherence tomography (OCT) is a label-free, non-destructive imaging modality with micron-scale spatial resolution and millimeter-scale depth penetration, and it provides 3D assessment of tissue with a few seconds of imaging time \cite{Huang1991Science}. 
However, conventional OCT can only visualize tissue morphology, and hence, it is not sufficient to investigate the metabolic changes associated with LD accumulation and inflammation during NAFLD pathogenesis.
For metabolism investigation, it is necessary to extend the contrast of OCT.

Among several contrast extensions of OCT, dynamic OCT is a recently developed technique used to visualize metabolic activity at the cellular or subcellular scale without any contrast agents \cite{apelian_dynamic_2016, munter_dynamic_2020, el-sadek_optical_2020, leung_imaging_2020, kurokawa_suite_2020, munter_microscopic_2021}.
Rapid time-sequential OCT acquisition and subsequent time-spectroscopic and/or statistical analysis contrasts the intracellular motion which is closely associated with tissue metabolism.
These methods has been successfully implemented in time-domain full-field \cite{apelian_dynamic_2016} and Fourier-domain OCT systems \cite{munter_dynamic_2020, el-sadek_optical_2020, leung_imaging_2020, kurokawa_suite_2020, munter_microscopic_2021}.
We recently demonstrated a custom-made 3D scanning protocol that uses a standard-speed swept-source OCT system to visualize volumetric tomography of tissue dynamics within a few seconds \cite{el-sadek_three-dimensional_2021}.

In this paper, we demonstrate 3D metabolic imaging of NAFLD mouse model liver using dynamic OCT.
We employ our recently demonstrated 3D dynamic OCT method\cite{el-sadek_three-dimensional_2021}, which enables volumetric dynamics imaging without exogenous labeling using a standard-speed (50,000 A-lines/s) swept-source OCT system. 
The mouse model is a high-fat methionine and choline-deficient (MCD) diet model.
Two mice were fed the MCD diet for either 1 or 2 weeks to induce hepatic steatosis and NASH, the early stages of NAFLD.
Dynamic OCT reveals several unseen 3D structures not observed in conventional OCT.
The results and extensive discussion show that these structures correspond to the metabolism associated with LD accumulation and inflammation.

\section{Methods}
\label{sec:methods}

\subsection{Non-alcoholic fatty liver disease (NAFLD) model and study design}
\label{subsec:NAFLD_model}
Three mice (wild type C57BL/6) were used in this study.
For the investigation of NAFLD, two mice were fed a methionine and choline-deficient (MCD) diet (Oriental Yeast Co., Ltd., Japan) to induce non-alcoholic steatosis and steatohepatitis \cite{anstee_mouse_2006, takahashi_animal_2012}.  
One of the two mice was fed the MCD diet for 7 days (1-week model), and the other for 14 days (2-week model).
One other mouse was fed a normal diet as a control.
All were almost 1-year old at the commencement of the diets and were housed in normal conditions and kept on a 12-hour light/dark cycle during the above-mentioned feeding period.

OCT and histological imaging were performed \exvivo.
The mice were sacrificed by cervical dislocation and their livers were dissected.
The livers were immersed in Dulbecco's modified eagle medium (DMEM, Sigma-Aldrich, MA) supplemented with 20\% fetal bovine serum (Sigma-Aldrich, MA) and penicillin/streptomycin (Gibco, MA) immediately after the dissection to avoid quick death of the tissues.
The OCT measurement was performed 1 hour after the dissection.


All animal experiments were performed in accordance with the animal study guidelines of the University of Tsukuba.
All experimental protocols involving mice were approved by the Institutional Animal Care and Use Committee of the University of Tsukuba.

\subsection{Dynamic OCT imaging by logarithmic intensity variance (LIV)}
\subsubsection{OCT system}
\label{subsec:system_protocol}
The OCT measurement was performed with a custom-made swept-source Jones-matrix OCT (JM-OCT) device.
The JM-OCT uses a MEMS-based wavelength sweeping laser at the 1.3 \um wavelength band (AXP50124-8, Axun Technologies, MA) with a sweeping rate of 50 kHz.
The objective (LSM03, Thorlabs Inc., NJ) has an effective focal length of 36 mm and the effective NA of the probe optics is 0.048. 
It leads to a lateral resolution of 18.1 \um.
The depth resolution is 14 \um in tissue.
The probe power on the sample was 12 mW.
More detailed descriptions are available elsewhere\cite{li_three-dimensional_2017, miyazawa_polarization-sensitive_2019}. 

Although the system is polarization-sensitive, only polarization-insensitive scattering OCT intensity was used in this study.
The polarization-insensitive OCT is the average of absolute-squared intensities of the four OCT signals corresponding to the system's four polarization channels.

\subsubsection{Scan protocol}
Volumetric dynamic OCT was obtained by combining a repeating raster scan protocol and a signal processing so-called logarithmic intensity variance (LIV) \cite{el-sadek_three-dimensional_2021}.
This method enables three-dimensional dynamics imaging within a few tens of seconds with the standard speed (50 kHz) OCT device.

For the repeating raster scan, the whole volumetric region (covering 3 mm $\times$ 3 mm, 5 mm $\times$ 5 mm, or 6 mm $\times$ 6 mm lateral region) is divided into four blocks where each block covers 3 mm $\times$ 0.75 mm, 5 mm $\times$ 1.25 mm, or 6 mm $\times$ 1.5 mm.
Each block is scanned by a raster scan protocol consisting of 32 frames along the fast-scan direction times 512 A-lines/frame. 
This raster scan is repeated 16 times, so each position on the sample is scanned 16 times, with an inter-frame-interval of 409.6 ms.
At each location, the time separation between the first and last frames is 6.55 s. 
This long frame separation results in high-contrast dynamic OCT, even with a small number of frames (Section 5.3 of Ref. \cite{el-sadek_optical_2020}).

Each block is then sequentially measured using this protocol to obtain a dynamic OCT volume.
This volume consists of 2,048 frames in total and the lateral volume size is 512 A-lines/frame times 128 frame locations.
The total acquisition time for a volume is 26.2 s.

\subsubsection{Signal processing for dynamics imaging}
\label{subsubsection:Signal processing}
The tissue dynamics is revealed by computing the LIV at each point in the volume.
Here, we model linear OCT intensity as $I(x,z,t_i) = I_D(x,z,t_i) I_S(x,z)$ where $I_D(x,z,t_i)$ and $I_S(x,z)$ are a dynamic and static components of the OCT signal, $x$ and $z$ are the lateral and axial positions, and $t_i$ is the acquisition time point of the $i$-th frame.
The dB-scaled OCT signal becomes $I_{dB}(x,z,t_i) = 10\log I_D(x,z,t_i) + 10 \log I_S(x,z)$.
The LIV is defined as the time variance of the logarithmic (dB)-scale OCT intensity signal \cite{el-sadek_optical_2020} as
\begin{align}\begin{split}
	\label{eq:LIV}
	\mathrm{LIV}(x,z)
	&= 	\frac{1}{N}\sum_{i = 0}^{N-1}
	\left[I_{dB}(x,z,t_i) - \ave{I_{dB}(x,z)}{} \right]^2,\\
	&= \frac{1}{N}\sum_{i = 0}^{N-1}
	\left[10\log I_D(x,z,t_i) - \ave{10\log I_D(x,z,t_i)}{} \right]^2,
\end{split}\end{align}
where $N$ is the number of frames, 16 in the present case.
$\ave{\quad}{}$ represents averaging over the $N$ frames.
Equation (\ref{eq:LIV}) suggests that LIV is sensitive to the fluctuation magnitude of the dynamic component, but is insensitive to the static component. 

In later sections, the LIV images are shown as pseudo-color composite image as described in Section 2.3 of \cite{el-sadek_optical_2020}.
In this composite image, the hue is the LIV, the pixel brightness is the averaged dB-scale OCT intensity over $N$ frames, and the saturation is set to 1 (full saturation) for all pixels.
The phase of the hue was selected to show regions of high and low dynamics in green and red, respectively.

\subsubsection{Slab projection image}
\label{subsec:volrender}
To generate an \enface slab projection image, the tissue surface was first identified from the dB-scale OCT intensity image by applying the surface segmentation algorithm as described in Ref. \cite{miyazawa_polarization-sensitive_2019}.
Here the OCT image is the average of 16 frames.
The region from the tissue surface to a depth of 100-pixel (724 \um) was extracted.
This process was performed frame-by-frame for the entire 3D dataset to form a slab.  
Slab projection images of OCT and LIV were obtained by averaging the OCT intensity and LIV. 

\subsection{Birefringence and degree-of-polarization-uniformity computation}
\label{subsec:BR-DOPU_computation}

The depth-resolved birefringence was computed using a local Jones matrix analysis method \cite{makita_generalized_2010} and maximum a-posteriori (MAP) estimator \cite{kasaragod_noise_2017}. 
The local Jones matrix analysis was applied with 8-pixel depth separation (57.9 \um) and the MAP estimation was performed with a spatial kernel of 2 pixels (23.4 \um, lateral) $\times$ 2 pixels (14.5 \um, depth). 

Degree of polarization uniformity (DOPU) was computed over a spatial kernel of 3 $\times$ 3 pixels with Makita's noise correction \cite{makita_degree_2014}.

\subsection{Histology}
\label{subsec:hist_imag}
Hematoxylin and eosin (H\&E) staining histologies were obtained after the OCT measurement. 
To perform H\&E staining, liver samples were fixed using 4\% paraformaldehyde (PFA) (Nacalai Tesque, Kyoto, Japan).
The samples were then embedded in paraffin, and tissue sections of 12 \um thickness were prepared.
The sections were imaged under a microscope (BZ-X710, Keyence Corp., Osaka, Japan) with an 10$\times$ objective (NA = 0.3).

\section{Results}
\subsection{Normal-diet induced mouse liver}
\label{subsec:normal_diet}
\begin{figure}
	\centering
	\includegraphics{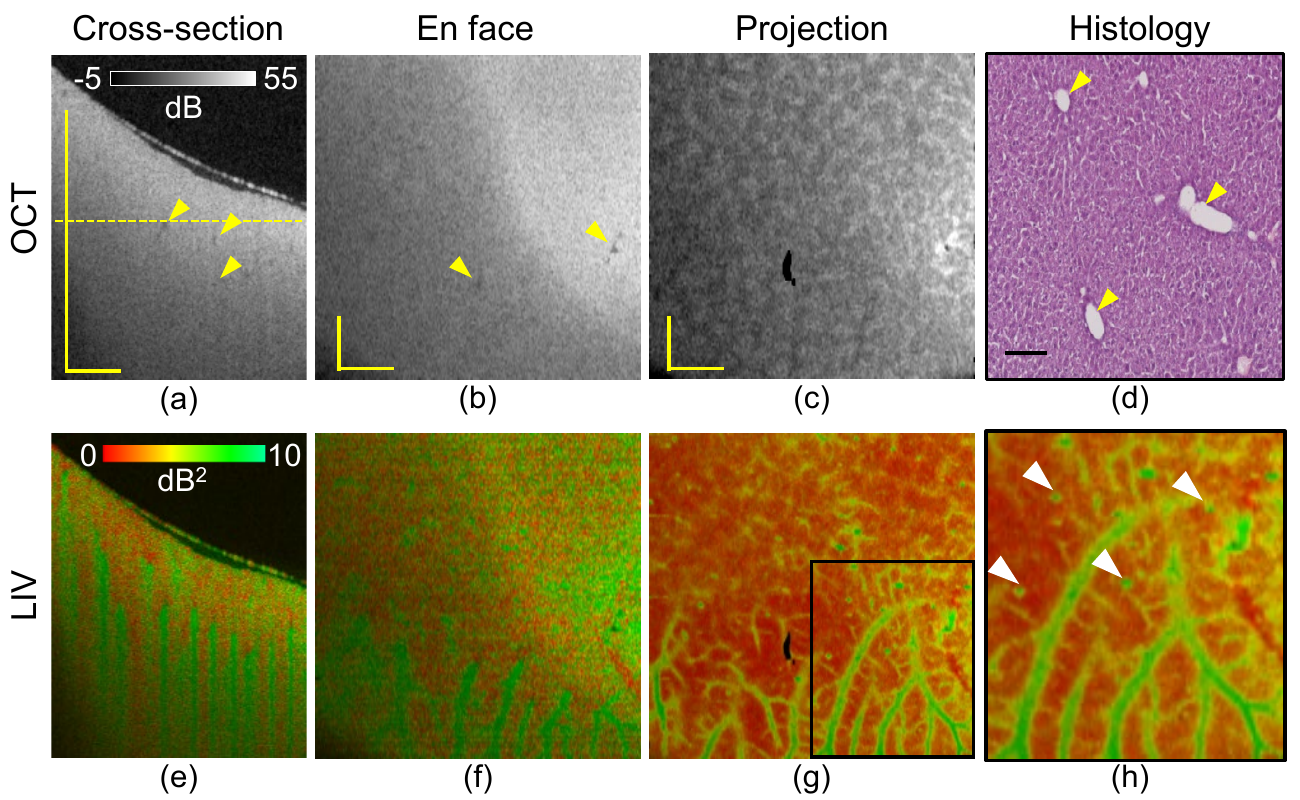}
	\caption{Dynamics imaging of a normal-diet induced mouse liver. 
		Cross-sections of (a) scattering OCT and (e) LIV images. 
		(b, f) \Enface slices of the OCT and LIV images at the depth location indicated by the horizontal line in (a). 
		(c, g) Slab projection images of OCT and LIV. 
		(d) H\&E staining histological micrograph.
		(h) Magnified image of the LIV projection image at the region marked by the rectangular box in (g). 	
		All scale bars indicate 1 mm.}
	\label{fig:normal_diet}
\end{figure}

Figure \ref{fig:normal_diet} summarizes the 3D imaging results of the normal-diet-induced mouse liver.
The depth location of the \enface slice is indicated by the horizontal line in the cross-sectional OCT image [\fig \ref{fig:normal_diet}(a)].
The thin layer at the top in \fig \ref{fig:normal_diet}(a) is the surface of the cultured medium. 

In the OCT intensity cross-section and the \enface slice [\fig \ref{fig:normal_diet}(a) and (b)], the tissue appears homogeneous except for some small hypo-scattering spots (yellow arrows).
Similar appearance was found in our previous study \cite{mukherjee_label-free_2021} and are known to be blood vessels.
A tessellated texture appears in the slab projection of OCT [\fig \ref{fig:normal_diet}(c)].  

Unlike the homogeneous appearance of the OCT image, the LIV shows several high-contrast structures.
In the deep tissue region of the cross-sectional image [\fig \ref{fig:normal_diet}(e)], several vertical stripes with high LIV (green) are observed.
Corresponding high-LIV structures are also found in the \enface image [\fig \ref{fig:normal_diet}(f), around the bottom]. 
Since high LIV indicates high temporal fluctuation of the OCT signal, high intracellular motion would be occurring in these high LIV structures.

The slab projection of the LIV [\fig \ref{fig:normal_diet}(g)] reveals that these high LIV regions form vessel-like structures.
In the magnified region of the slab projection [\fig \ref{fig:normal_diet}(h)], several small dots can also be seen around the vessel-like structures (white arrowheads).
The vessel-like structure and the small dots are expected to be perivascular tissue and lipid droplets, as we discuss later in Sections \ref{subsec:Metabolism_normal-diet} and \ref{subsec:LD_motion}.
It is noteworthy that the patterns appeared in the OCT intensity and LIV are totally uncorrelated.

Figure \ref{fig:normal_diet}(d) shows representative H\&E histology of the same sample.
Normal lobular patterns can be observed.
The void structures are central and portal veins (yellow arrowheads).

\subsection{1-week MCD-diet-induced NAFLD model}
\label{subsec:1-weekModel}
\begin{figure}
	\centering
	\includegraphics{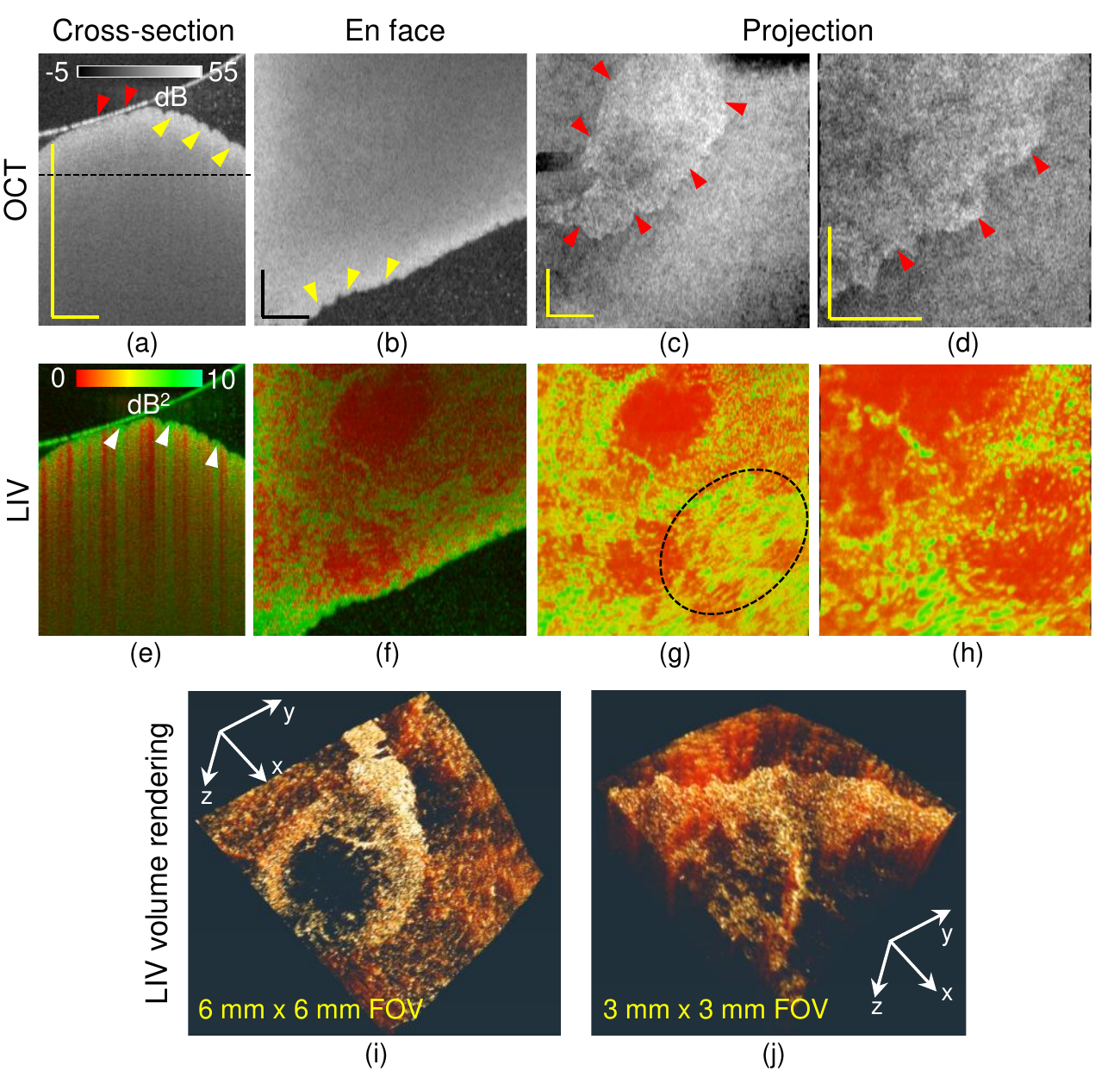}
	\caption{Imaging results of a 1-week MCD diet induced NAFLD model. 
		Cross-sections of (a) scattering OCT and (e) LIV images. 
		(b, f) \Enface slices of the OCT and LIV images at the depth location indicated by the horizontal line in (a), where the transversal field of view (FOV) is 6 mm $\times$ 6 mm.
		(c, g) OCT and LIV slab projections and (i) volume rendering of the LIV tomography.
		(d, h) OCT and LIV slab projections and (j) LIV volume rendering of narrow FOV (3 mm $\times$ 3 mm) measurement.
		All scale bars indicate 1 mm.}
	\label{fig:1wkmodel}
\end{figure}

Figure \ref{fig:1wkmodel} shows the tomographies of a 1-week MCD-diet-induced NAFLD model.
Cross-sectional and \enface slices of the OCT intensity [\fig \ref{fig:1wkmodel}(a) and (b)] reveal that the surface is wavy (yellow arrowheads).
Such unevenness was not observed in the normal-diet liver [\fig \ref{fig:normal_diet}].
The tissue is mostly homogeneous except at the surface [\fig \ref{fig:1wkmodel}(a) and (b)].
In the cross-section, the thin, hyper-reflective layer above the tissue surface is the surface of the cultured medium, and the dots in the medium are floaters.
Although the slab projections [\fig \ref{fig:1wkmodel}(c) and (d)] show a clear domain structure, it is an artifact. 
Namely, in one domain, the tissue was totally soaked in the cultured medium [yellow arrowheads, \fig \ref{fig:1wkmodel}(a)], while in the other domain, the tissue surface contacted the surface of the medium [red arrowheads, \fig \ref{fig:1wkmodel}(a)]. 
Hence, the segmented surfaces of two domains are not the same types of interface.
This inconsistency of the segmentation created the pseudo-domain border [red arrowheads, \fig \ref{fig:1wkmodel}(c) and (d)].

The cross-sectional LIV image [arrowheads, \fig \ref{fig:1wkmodel}(e)] reveals that the tissue surface region exhibits high temporal fluctuation (green).
Several vertical stripes of high LIV (green) are also observed. 
Since its appearance is similar to the projection artifact of OCT angiography \cite{spaide_image_2015}, they are expected to be the projection artifacts of LIV.

In the \enface slice of the LIV image [\fig \ref{fig:1wkmodel}(f)], the regions of high-LIV signals form macroscopic, ring-shaped structure where the high LIV region shows a granular appearance.
Here the field of view (FOV) is 6 mm $\times$ 6 mm.
The slab projections of the LIV [\fig \ref{fig:1wkmodel}(g)] shows a similar pattern with the \enface slice.
However, it also reveals some structural flows (dashed circle).
Similar appearances are also found in the slab projection of a smaller-FOV measurement [\fig \ref{fig:1wkmodel}(h)] (3 mm $\times$ 3 mm).
The volume renderings of the LIV [\fig \ref{fig:1wkmodel}(i) and (j)] show the 3D morphology of the ring-shaped, high-LIV structures.
For visualizing the volume rendering images, a 3D visualization software Amira (Thermo Fisher Scientific, MA, USA) was used. 
As we will discuss later in Section \ref{subsec:LD_motion}, these high-LIV signals including the small particles are caused by the motion of lipid droplets.

Note that, histology was not successfully obtained from this sample because the liver was too oily and it adhered to the blades during the cryosection procedure.

\subsection{2-week MCD-diet-induced NAFLD model}
\label{subsec:2-weekModel}

\begin{figure}
	\centering
	\includegraphics{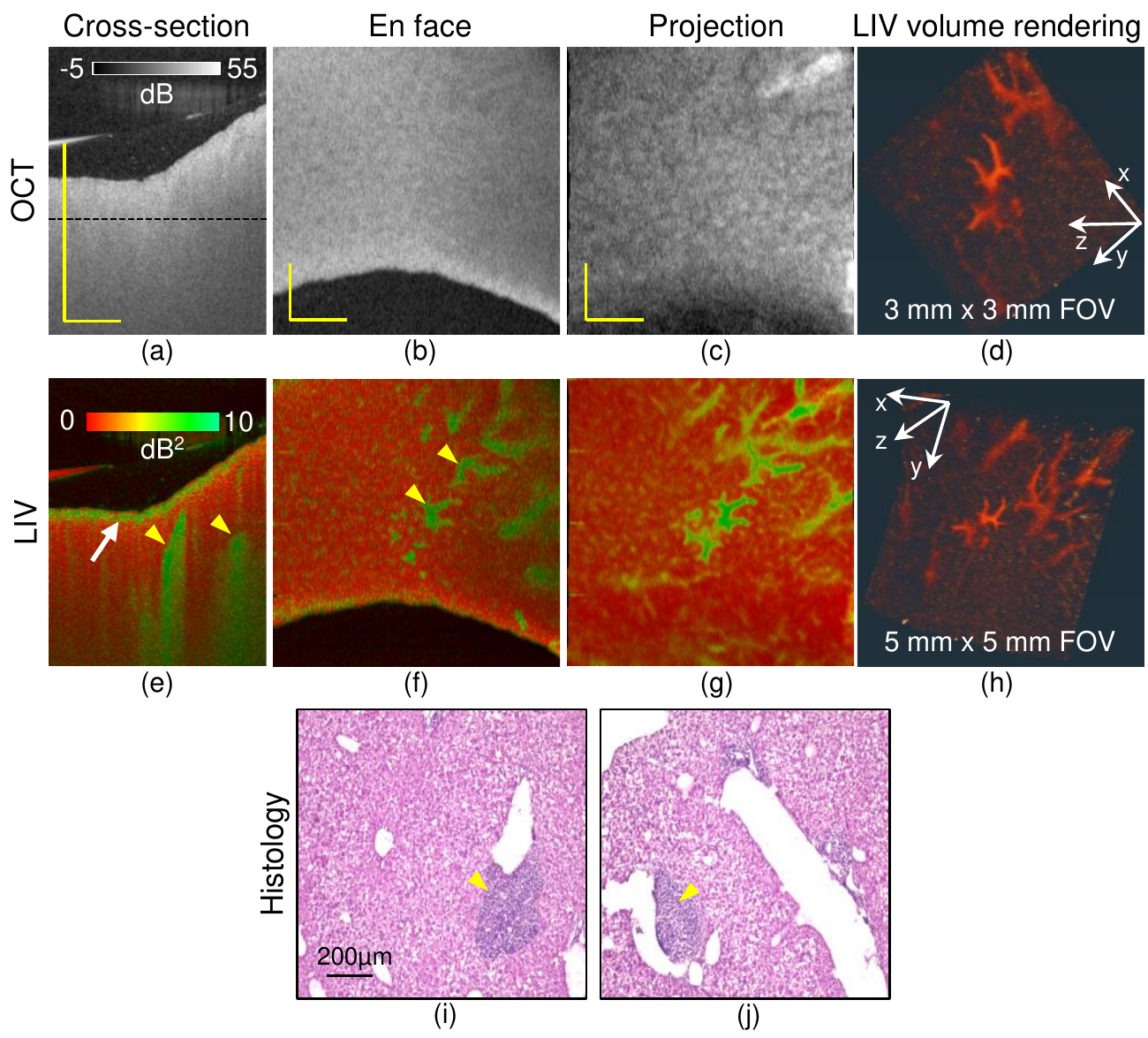}
	\caption{%
		The results of 2-week MCD diet induced NAFLD model liver. 
		The first to the third columns show the cross-sectional images (a, e), the \enface slices (b, f), and the slab projections (c, g) of OCT intensity (first row) and LIV (second row) respectively. 
		The fourth column (d, h) shows the volume renderings of LIV with two different FOVs.
		The last row (i, j) shows the H\&E stained histological micrographs. 
		The scale bars of OCT intensity and LIV represent 1 mm and that in histology indicate 200 \um.}
	\label{fig:2wkmodel}
\end{figure}

Figure \ref{fig:2wkmodel} shows OCT, LIV, and histological images of the 2-week NAFLD mouse liver.
The cross-section and \enface slice of the OCT intensity [\fig \ref{fig:2wkmodel}(a) and (b)] appear similar to those of the normal-diet mouse [\fig \ref{fig:normal_diet}(a) and (b)]. 
The horizontal black line in the cross-section indicates the depth location of the \enface slice.
The slab OCT projection [\fig \ref{fig:2wkmodel}(c)] shows a granular hyper-scattering pattern, which is not identical to the tessellated pattern of the normal-diet liver  [\fig \ref{fig:normal_diet}(c)]. 

The LIV cross-section [\fig \ref{fig:2wkmodel}(e)] exhibits a high-LIV (green) layer near the surface (white arrow), but unlike the 1-week MCD model [Fig. \ref{fig:1wkmodel}(e)], this high-LIV layer is slightly below the surface.
The cross-sectional image also exhibits vertical high-LIV (green) stripes (yellow arrowheads), but there are fewer stripes than in 1-week model, and they appear in a deeper region.
In the \enface slice [\fig \ref{fig:2wkmodel}(f)], this high-LIV signal shows fragmented, vessel-like structures (yellow arrowheads).
The same structures are more evident in the slab projection [\fig \ref{fig:2wkmodel}(g)] and the volume rendering [\fig \ref{fig:2wkmodel}(h)].
The volume rendering of a narrower-FOV measurement [3 mm $\times$ 3 mm, \fig \ref{fig:2wkmodel}(d)] also shows the same structure.
It is noteworthy that this structure is not revealed in the OCT intensity images.

The H\&E histology micrographs of the same sample [\fig \ref{fig:2wkmodel}(i) and (j)] exhibit deep purple regions around the large vessels (yellow arrowheads), which may suggest inflammatory cell infiltration as we will discuss in Section \ref{subsec:Inflam_cells_metabolism}.
The normal lobular pattern seen in the normal-diet liver cannot be found in the present liver.
It may indicate hepatocellular ballooning as we will discuss in Section \ref{subsec:LD_motion}.

\section{Discussion}
\label{sec:Discussion}

\subsection{Vessel-like structures in the normal-diet liver: periportal and perivenous zones}
\label{subsec:Metabolism_normal-diet}
In the normal-diet liver, LIV allowed us to visualize 3D vessel-like structures [\fig \ref{fig:normal_diet}].
Although vascular structure was not clear in the OCT intensity in this study [\fig \ref{fig:normal_diet}(b) and (c)], our previous study of normal mouse liver suggests that the high-LIV locates just beneath the vascular lumen \cite{mukherjee_label-free_2021}.
Accordingly, the high-LIV vessel-like structures in this study also colocate with the tissue beneath or around vessels.

Here, we suspect that the high-LIV vessel-like structures correspond to high metabolic activity in the periportal/perivenous zone.
The liver is an organ that is responsible for whole-body homeostasis \cite{robinson_liver_2016}, and hence, it is highly metabolically active in general.
Liver tissue has multiple functional zones along the hepatic sinusoid, from the central vein to the portal vein\cite{birchmeier_orchestrating_2016, kusminski_new_2018}.
In the periportal zone, hepatic lobules are involved in oxidative energy metabolism, gluconeogenesis, and urea synthesis \cite{ kietzmann_metabolic_2017, kang_metabolic_2018}, and hence it is metabolically active.
In the perivenous zone (around the central vein), hepatocytes are activated by the metabolic pathways of glycolysis and liposynthesis \cite{kietzmann_metabolic_2017, kang_metabolic_2018}, and hence it is also metabolically active.
Namely, the tissues around the portal and central veins are metabolically active.
Appelian \etal showed that the contrast source of their dynamic OCT method, which is based on time-frequency analysis of OCT, is ATP-consuming motion \cite{apelian_dynamic_2016}.
Hence, the ATP-consuming motion in the metabolically-active periportal and perivenous tissues might be the source of the high-LIV signals of the vessel-like structures.

It is noteworthy that these vessel-like structures are hard to observe with other modalities.
As we have mentioned, the OCT intensity image in the current study does not clearly visualize them.
Although it was visible in our previous OCT intensity imaging \cite{mukherjee_label-free_2021}, the same study also showed that the OCT intensity does not always visualize the vessel lumens.
In H\&E histological micrographs [\fig \ref{fig:normal_diet}(d)], the cross-sections of the vessels were observed.
However, the micrographs did not clearly distinguish between the periportal and perivenous zones.
Fluorescence microscopy also cannot visualize these metabolically active vessel-like structures with millimeter thickness in 3D, because it can only penetrate a few hundred micrometers from the sample surface.

\subsection{Lipid droplets in LIV images}
\label{subsec:LD_motion}

LDs are lipid-storing cellular organelles comprising of a lipid core surrounded by a phospholipid monolayer and associated proteins.
They are key regulators of cellular metabolism \cite{kumar_natarajan_structure_2017, mashek_hepatic_2021}. 
A high-fat diet, such as the MCD diet, results in the accumulation of excessive fat in the liver.
It leads to hepatic steatosis, an initial stage of NAFLD \cite{angulo_nonalcoholic_2002}.
Hence, hepatic steatosis exhibits the accumulation of LD \cite{digel_lipid_2010}.

LDs are the lipid source for energy metabolism, and their movement is associated with nutrient transfer from one point to another \cite{welte_fat_2009}.
Hence, LDs exhibit highly dynamic 3D motion to regulate intracellular lipid storage \cite{martin_lipid_2006, welte_fat_2009}.
It may suggest that LDs exhibit high LIV signal. 

With the continuous feeding of the high-fat MCD diet, the LD diameter can be gradually increased \cite{moon_intravital_2020}.
Moon \etal observed the formation of a small number of hepatic LDs also in a normal-diet mouse, but with smaller diameter of around 1 \um \cite{moon_intravital_2020}. 
Namely, LDs can be observed both in normal-diet and MCD-diet mice, but its diameter and the number are larger for the latter mice.

In the normal-diet mouse liver [\fig \ref{fig:normal_diet}], small dots of high LIV were found around the vessel-like structures, which are suspected to be the periportal and perivenous tissues discussed in Section \ref{subsec:Metabolism_normal-diet}.
Because this mouse was fat even though it was fed a normal diet, some LDs may have formed in its hepatocytes.
Moon \etal showed fluorescent image of LDs in a normal-diet mouse liver \cite{moon_intravital_2020}, and our LIV appearance resembles their fluorescent image.
Hence, the small high-LIV dots might be LDs.

In the 1-week MCD model (\fig \ref{fig:1wkmodel}), the LIV signal forms a macroscopic ring shape.
Because this high-LIV ring is a collection of small high-LIV dots, it might also correspond to a collection of LDs.
It is also noteworthy that these high-LIV dots exhibit tail artifacts [arrowheads in \fig \ref{fig:1wkmodel}(e)], which may indicate that these LDs are highly motile.

In the 2-week MCD model, a thin layer of high LIV was observed near the tissue surface [\fig \ref{fig:2wkmodel}(e)].
Because the suspected LDs of the 1-week MCD model are also located around the surface, we suspect that this high-LIV layer in the 2-week model also represents LDs.
The histologies [\fig \ref{fig:2wkmodel}(i) and (j)] showed enlarged and ballooned hepatocytes.
Several previous studies of 2-week MCD models showed similar histology and accumulation of LDs \cite{overi_hepatocyte_2020, kleiner_histology_2016, hall_lipid_2017}, which also also supports our interpretation of this high-LIV appearance.

One difference between the 2-week and 1-week MCD models is that the latter exhibits tail artifacts in the LIV image [arrowheads in \fig \ref{fig:1wkmodel}(e)], while the former does not.
Moon \etal \cite{moon_intravital_2020} showed that the size of LDs increases with continuous feeding with the MCD diet.
So, the LDs in the 2-week model might be larger than those in the 1-week model.
This difference in size can be associated with the different appearances of the LIV and its tail artifacts, but further investigation is required.

As we have discussed, LDs play a key role in liver metabolism.
Hence, LIV can contribute to liver metabolism studies.

\subsection{High-LIV structure as an indicator of inflammation of NASH}
\label{subsec:Inflam_cells_metabolism}

Several fragmented vessel-like structures are visible in the LIV projection [\fig \ref{fig:2wkmodel}(g)] and volume rendering [\fig \ref{fig:2wkmodel}(d) and (h)] of the 2-week MCD-diet mouse liver.
Based on their location and size, the vessel-like structures may correspond to the purple clusters seen in the histology around relatively large vessels [\fig \ref{fig:2wkmodel}(i) and (j)].
Previous histological investigations of 2-week MCD-diet mouse livers suggest that these are inflammatory cell regions \cite{takahashi_histopathology_2014, zhang_cxc_2016}.

Inflammatory immune cells infiltrate in NASH liver \cite{gao_inflammation_2016} including Kupffer cells, neutrophils, natural killer cells, and dendritic cells \cite{ganz_immune_2013}.
They produce various pro-inflammatory cytokines, such as tumor necrosis factor (TNF)-$\alpha$ and interleukin-6 (IL-6) \cite{silvana_inflammatory_2018, lopetuso_harmful_2018}, which activate tissue metabolism.
For example, TNF‑$\alpha$, which stimulates liver steatosis, increases serum triglyceride levels, hence supplying energy to the tissue \cite{yang_roles_2019}. 
Similarly, IL‑6 is an important mediator of fatty acid metabolism\cite{vida_chronic_2015}.

This link between inflammatory cells and metabolism may explain the high-LIV appearance of the fragmented vessel-like structures, which would allow LIV to be used as an imaging biomarker of NASH. 

\subsection{Birefringence contrast investigation during NAFLD pathogenesis}
\label{subsec:BR_contrast}
The inflammatory immune cells that infiltrate during NASH development activate the activation chain of pro-inflammatory cytokines, hepatic stellate cells, and myofibroblasts \cite{kietzmann_metabolic_2017}.
Myofibroblasts produce excessive amounts of extracellular matrix components and lead to hepatic fibrosis, followed by cirrhosis, which is the almost end-stage of NAFLD.
Currently, histology is the standard technique for accurately identifying liver fibrosis, but it has several limitations including invasiveness, sampling error, labor-intensive staining, and long imaging time. 

As mentioned in Section \ref{subsec:system_protocol}, we employed a polarization-sensitive OCT device (JM-OCT) in this study, but we did not use its polarization-sensitive imaging function.
Among the polarization-sensitive JM-OCT images, birefringence is known to be sensitive to fibrotic tissue \cite{fukuda_comparison_2018}.
Therefore, future studies with the polarization-sensitive function of JM-OCT might be helpful for understanding the fibrosis process and pathogenesis of NAFLD.

\subsection{Comparison of the polarization contrasts among NAFLD stages}
\label{subsec:polarization_contrasts}

\begin{figure}
	\centering
	\includegraphics{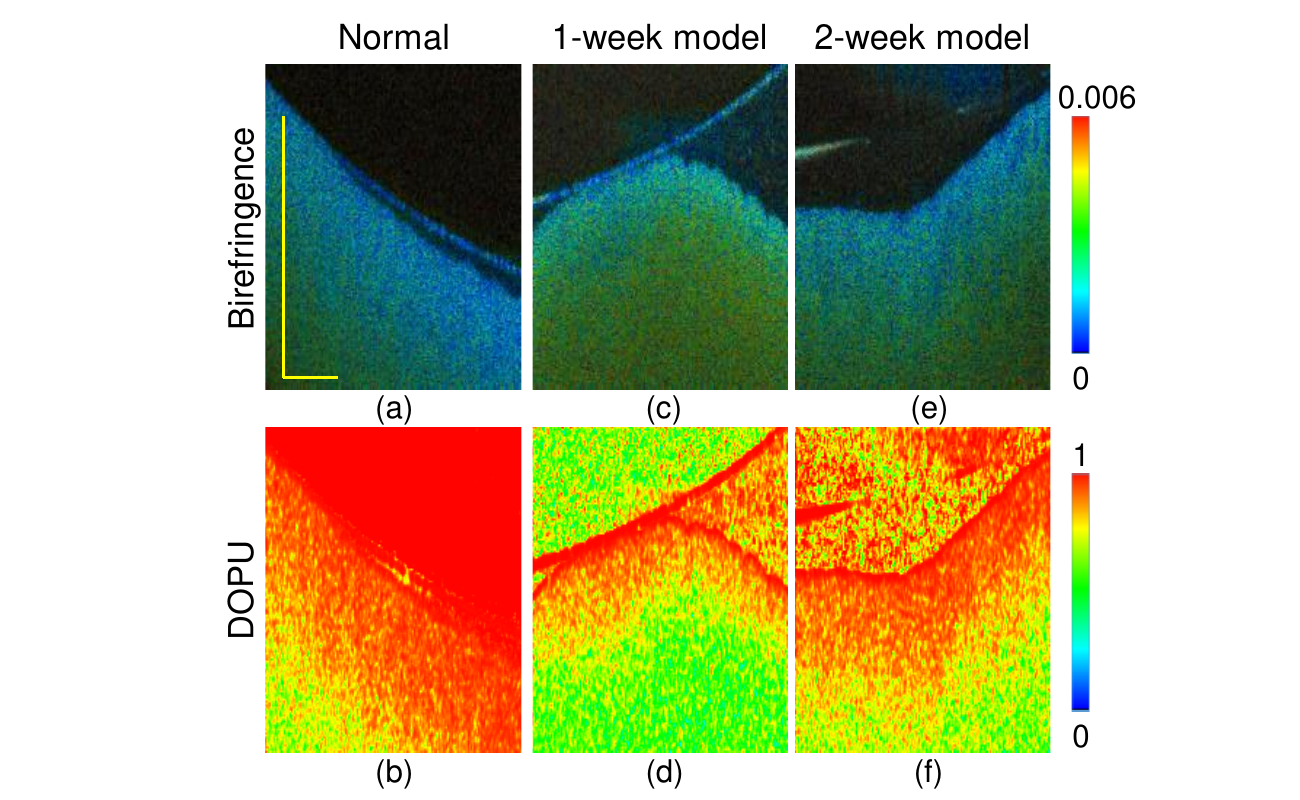}
	\caption{%
		Cross-sectional birefringence (a, c, e) and DOPU (b, d, f) images of the normal liver (first column), the 1-week NAFLD model (second column), and the 2-week NAFLD model (third column), respectively.  
		The scale bars represent 1 mm.}
	\label{fig:BR_DOPU}
\end{figure}

Since the OCT device used in this study is a JM-OCT \cite{li_three-dimensional_2017, miyazawa_polarization-sensitive_2019}, polarization sensitive images, i.e., birefringence and DOPU images, can be obtained in addition to the polarization-insensitive OCT image. 
However, the polarization-sensitive images were hard to interpret. 
Hence, we omitted these images from the main part of this study. 
For the readers' interest, here we discuss about the polarization-sensitive images, although it is inconclusive.

Figure \ref{fig:BR_DOPU} are the examples of polarization-sensitive images of normal, 1-week and 2-week NAFLD model livers.
Figure \ref{fig:BR_DOPU}(a) and (b) are cross-sectional birefringence and DOPU of normal liver, respectively. 
The birefringence changes along the depth from relatively low (blue) to moderately high (green). 
Similarly, DOPU decreases gradually (red to green) along the depth. 
In the 1-week NAFLD model [Fig.\@ \ref{fig:BR_DOPU}(c) and (d)], the birefringence appears to be moderately high (green), while the DOPU is low (green). 
In the 2-week NAFLD model [Fig.\@ \ref{fig:BR_DOPU}(e) and (f)], similar appearances of birefringence and DOPU were observed to those of the normal liver. 
Namely, the birefringence is low and DOPU is high at the region close to the surface, and they become moderately high (birefringence) and low (DOPU) in the deep region. 

The interpretation of the high birefringence at low DOPU regions is an open issue. 
In our JM-OCT, both birefringence and DOPU are computed with a spatial kernel; 2 $\times$ 2 pixels, i.e., 14.5 \um (depth) $\times$ 23.4 \um (lateral) for birefringence, and 3 $\times$ 3 pixels, i.e., 21.7 \um (depth) $\times$ 35.2 \um (lateral) for DOPU \cite{kasaragod_noise_2017, makita_degree_2014}. 
For the birefringence computation, the tissue is assumed to be homogeneous within the kernel. 
However, low DOPU appearance suggests that the tissue is not homogeneous in reality, and hence, the estimated birefringence is not reliable at this region. 
Conversely, if the true birefringence of the sample is high, Stokes vectors, which are used for DOPU computation, alter along the depth, and DOPU becomes low. 
Hence, low DOPU at a high birefringence region does not necessarily mean random polarization, which is the main interest of DOPU measurement.

\subsection{Comparison of dynamics imaging methods}
\label{subsec:dynamics_methods}
The dynamic-OCT methods visualize tissue dynamics by analyzing the temporal characteristics of time-sequential OCT data. 
Several variations of dynamic-methods have been developed. 
These methods can be categorized into two types.
Type-1 methods are methods which are sensitive to the magnitude of the signal fluctuation, while type-2 methods are sensitive to the speed of the fluctuation.
Examples of type-1 methods include standard deviation of time-sequential signals \cite{oldenburg_motility-_2013, oldenburg_inverse-power-law_2015, apelian_dynamic_2016}, cumulative-sum method \cite{scholler_motion_2019}, and LIV \cite{el-sadek_optical_2020}. 
The type-2 methods are further subdivided into two subtypes. 
One (type 2-1) is correlation decay analysis including dynamic scattering OCT \cite{lee_dynamic_2012}, inverse-power-law analysis \cite{oldenburg_inverse-power-law_2015}, exponential fitting of decorrelation curve \cite{kurokawa_suite_2020}, and OCT-correlation-decay-speed analysis \cite{el-sadek_optical_2020, el-sadek_three-dimensional_2021}. 
The other (type 2-2) is power spectrum analysis \cite{scholler_dynamic_2020, leung_imaging_2020, munter_dynamic_2020}.

Type-1 methods, including presently used LIV, provide a high-contrast image of dynamic structures.
However, its ability for quantitative assessment is limited. 
In contrast, type-2 methods provide more quantitative fingerprints of tissue dynamics. 
For example, Lee \etal used complex correlation analysis of time-sequential OCT signals to extract the velocity and diffusion coefficient of moving particles \cite{lee_dynamic_2012}. 
Through power-spectrum analysis of time-sequential OCT signals, Scholler \etal visualized several specific activities in retinal organoids \cite{scholler_dynamic_2020}.  

In the present study, we used only a type-1 method (LIV) to investigate the mouse livers. 
In order to obtain the 3D distribution of LIV with a standard-speed OCT, we acquired only 16 frames at a single location. 
It prevents us from performing the correlation decay analysis that requires around 30 frames as mentioned in Section 5.3.2 of \cite{el-sadek_optical_2020}.
For some less dynamic samples such as \invitro spheroids than the mouse liver, we can obtain 32 frames at a single location, and it enables correlation speed analysis \cite{el-sadek_three-dimensional_2021}. 
In the future, new scanning protocol and signal processing methods \cite{morishita_sparse_2022} might enable type-2 methods with a smaller number of time-sequential OCT frames.

\subsection{Projection artifact of LIV}
\label{subsec:projection_artifact}
\begin{figure}
	\centering
	\includegraphics{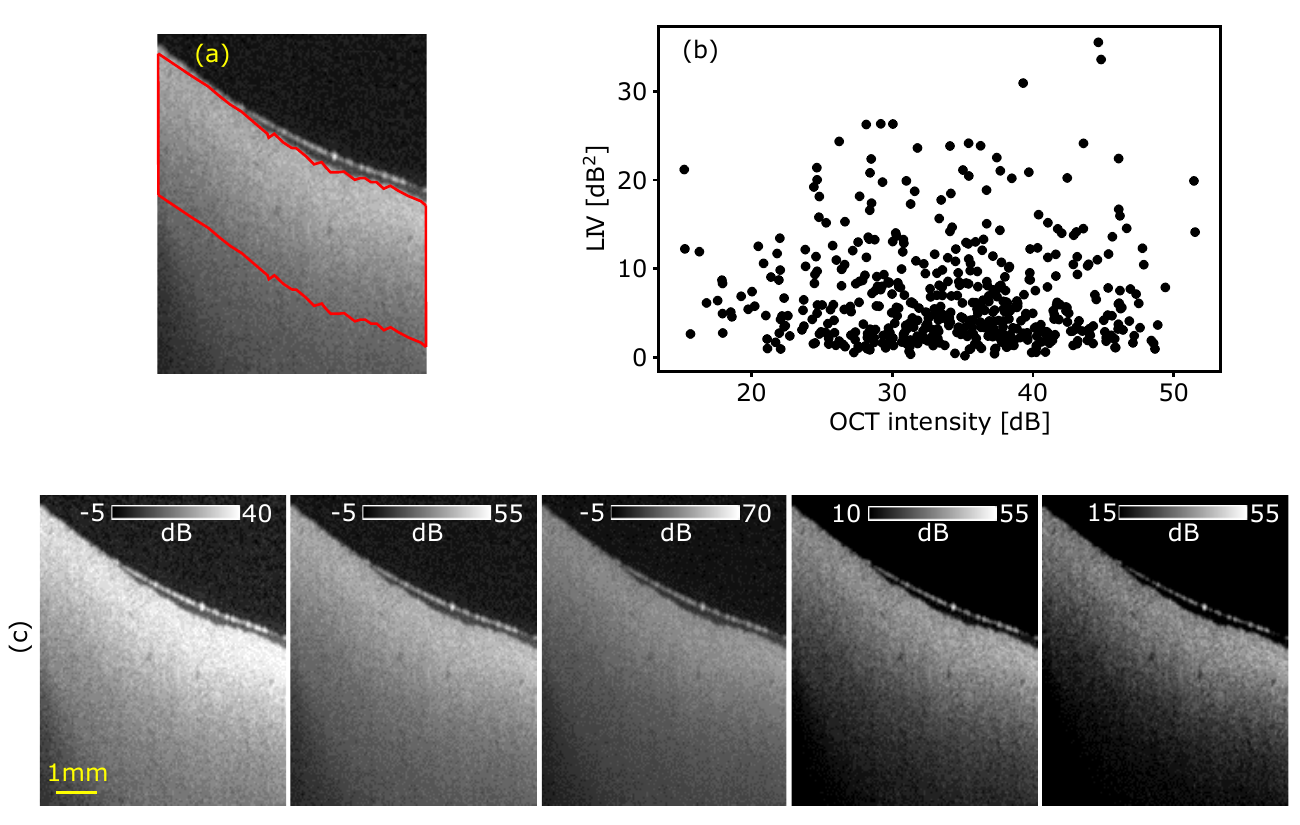}
	\caption{%
	Scatter plot between dB-scale OCT intensity and LIV (b) at the region delineated by red line in (a). 
	No evident correlation was found between the dB-scale OCT intensity and LIV.
	(c) The OCT intensity of Fig.\@ \ref{fig:normal_diet}(a) is displayed at different intensity ranges.
	The vertical stripe structure shown in the LIV image [Fig.\@ \ref{fig:normal_diet}(e)] is not visible in the intensity images.}
	\label{fig:artifact}
\end{figure}

The cross-sectional LIV images of normal [\fig \ref{fig:normal_diet}(e)] and 1-week model [\fig \ref{fig:1wkmodel}(e)] exhibited several vertical high-LIV signals. 
We hypothesize that this vertical LIV appearance is not associated with a real structure which could appear in the OCT intensity images or an artifact due to the static intensity OCT pattern, but a projection artifacts caused by high dynamics of superficial tissues similar to that of OCTA. 
This hypothesis can be supported by the following discussion.

According to the mathematical model (Section \ref{subsubsection:Signal processing}), LIV can be insensitive to the static component of the OCT intensity but sensitive mainly to the fluctuation magnitude of the dynamic component. 
To further validate this point, we did additional analysis on the data of \fig \ref{fig:normal_diet}(a). 
Figure \ref{fig:artifact}(b) shows a scatter plot between the dB-scale OCT intensity and the LIV, where each point represents a single pixel, and 500 pixels randomly selected in a tissue region [delineated by red line in Fig.\@ \ref{fig:artifact}(a)] were plotted.
As evident in the plot, the correlation between the OCT intensity and LIV was low (r = 0.15). 
Furthermore, the similar vertical structure to that of LIV cannot be seen in the OCT intensity image [Fig.\@ \ref{fig:artifact}(c)], where the OCT images are displayed with several intensity ranges in order to highlight possible low-contrast structures.

According to these results, we believe the vertical stripe of LIV is not an artifact caused by spatial distribution of OCT intensity.
It might be a future task to prove this stripe structure is a projection artifact similar to OCTA. 
This can be further investigated by using fluctuation speed analysis (type-2 methods discussed in Section \ref{subsec:dynamics_methods}).

It is noteworthy that the \enface LIV images, such as \fig \ref{fig:normal_diet}(g) and (h) and \fig \ref{fig:1wkmodel}(g) and (h) provide more interpretable structures, which can be directly related to metabolic regions, than the cross-sectional images. 
This may be due to the fact that \enface projection images are less disturbed by the projection artifact.

\section{Conclusion}
\label{sec:conclusion}
The metabolic activity of normal-diet, MCD-diet-induced 1-week, and 2-week NAFLD model mouse livers were investigated in 3D and label-free approach by volumetric dynamic OCT. 
In the normal-diet liver, highly dynamic vessel-like structures and several highly dynamic dots were observed.
The former might correspond to high metabolic activity in the periportal and perivenous zones, while the latter indicate accumulation of LDs.
In the 1-week NAFLD model, several vertical stripes of high-LIV signal were observed, which may indicate highly-motile LDs and their projection artifacts.
In the \enface and volume-rendering observations, the high-LIV regions form macroscopic, ring-shaped structure and reveal structural flows.
In the 2-week model, highly dynamic fragmented vessel-like structures were observed, which are believed to correspond to inflammatory cells around large vessels.

In conclusion, LIV imaging can be used to visualize the tissue metabolism associated with LD accumulation and inflammation without exogenous labels.
Hence, LIV can be used to detect these abnormalities of the NAFLD model.
In the future, LIV can be used for more precise quantification of tissue abnormality in NAFLD models, and hence will contribute to understanding of NAFLD pathogenesis.

\section*{Funding}
	Core Research for Evolutional Science and Technology (JPMJCR2105);
	Japan Science and Technology Agency (JPMJMI18G8);
	Austrian Science Fund (Schrödinger grant, J4460);
	Japan Society for the Promotion of Science (18H01893, 21H01836).
			
\section*{Acknowledgments}
	Although this research was funded solely by the agencies listed, the project is related to a joint research project between Yokogawa Electric Corp. and the University of Tsukuba. 
	The authors greatly appreciate fruitful technical discussions with Hiroyuki Sangu (Yokogawa), Atsushi Kubota and Renzo Ikeda (Skytechnology), Akihiro Shitoh, and Yuichi Inoue (Optosigma), Masato Takaya (Tatsuta), and Naoki Fukutake (Nikon).
	
\section*{Disclosures} 
	Mukherjee, El-Sadek, Zhu, Morishita, Lichtenegger: Yokogawa Electric Corp. (F), Sky Technology (F), Nikon (F), Kao Corp. (F), Topcon (F). 
	Makita, Yasuno: Yokogawa Electric Corp. (F), Sky Technology (F), Nikon (F), Kao Corp. (F), Topcon (F), Tomey Corp (P). 
	Okada: None.
	Miyazawa: Sky Technology (E).
	Fukuda, Lukmanto, Yamashita: None.
	Oshika: Topcon (F), Tomey Corp (F). 
	\smallskip
	
\section*{Data Availability Statement} 
	The data that support the findings of this study are available from the corresponding author upon reasonable request.
	
\bibliography{NAFLD_2021}

\end{document}